\documentclass[conference]{IEEEtran}

\IEEEoverridecommandlockouts
\usepackage{cite}
\usepackage{amsmath,amssymb,amsfonts} \usepackage{algorithmic} \usepackage{graphicx}
\usepackage{textcomp} \usepackage[table,xcdraw]{xcolor} \usepackage{siunitx}
\usepackage{multirow}
\usepackage{booktabs}
\usepackage{tabularx}
\usepackage{url}
\def\BibTeX{{\rm B\kern-.05em{\sc i\kern-.025em b}\kern-.08em
    T\kern-.1667em\lower.7ex\hbox{E}\kern-.125emX}}

\DeclareSIUnit\pixel{px}

\begin{document}

\title{Affective Air Quality Dataset: Personal Chemical Emissions from Emotional Videos}

\author{
\IEEEauthorblockN{Jas Brooks}
\IEEEauthorblockA{\textit{Microsoft Research}\\
Cambridge, USA \\
jasb@mit.edu}
\and
\IEEEauthorblockN{Javier Hernandez}
\IEEEauthorblockA{\textit{Microsoft Research}\\
Cambridge, USA \\
javierh@microsoft.com}
\and
\IEEEauthorblockN{Mary Czerwinski}
\IEEEauthorblockA{\textit{Microsoft Research}\\
Redmond, USA \\
marycz@microsoft.com}
\and
\IEEEauthorblockN{Judith Amores}
\IEEEauthorblockA{\textit{Microsoft Research}\\
Cambridge, USA \\
judithamores@microsoft.com}
}

\maketitle

\begin{abstract}
Inspired by the role of chemosignals in conveying emotional states, this paper introduces the Affective Air Quality (AAQ) dataset, a novel dataset collected to explore the potential of volatile odor compound and gas sensor data for non-contact emotion detection. This dataset bridges the gap between the realms of breath \& body odor emission (personal chemical emissions) analysis and established practices in affective computing. Comprising 4-channel gas sensor data from 23 participants at two distances from the body (wearable and desktop), alongside emotional ratings elicited by targeted movie clips, the dataset encapsulates initial groundwork to analyze the correlation between personal chemical emissions and varied emotional responses. The AAQ dataset also provides insights drawn from exit interviews, thereby painting a holistic picture of perceptions regarding air quality monitoring and its implications for privacy. By offering this dataset alongside preliminary attempts at emotion recognition models based on it to the broader research community, we seek to advance the development of odor-based affect recognition models that prioritize user privacy and comfort.
\end{abstract}


\section{Introduction}

Recent studies have illuminated the critical role of chemosignals in conveying emotional states through alterations in bodily emissions \cite{camarinha-matos_real_2020, de_groot_human_2017, de_groot_sniff_2015, calvi_scent_2020}. This phenomenon underpins the potential of using gas sensing technologies to detect emotional states as a potential new modality for existing affective computing systems. Understanding and accurately detecting human emotions presents a challenge compounded by privacy concerns with audiovisual methods \cite{alharbi_mask_2019} and user discomfort with contact-based sensors. Motivated to addressing these issues, this paper introduces the \emph{Affective Air Quality} (AAQ) Dataset, which endeavors to explore the potential of gas sensors as a novel, non-contact modality for emotion detection in a personal manner. By analyzing changes in breath and body emission composition, the AAQ Dataset seeks to offer an unobtrusive alternative to traditional affective computing approaches, mitigating privacy and comfort concerns associated with methods that rely on audio-visual input or direct physical contact. We describe the \emph{first-of-its-kind}, open-access dataset that captures real-time volatile odor compound (VOC) and gas sensor data (personal chemical emissions) while participants experience different emotional responses elicited by movie clips in a controlled setting. Sensor data are obtained by two custom-made devices (a wearable and a standalone), that use off-the-shelf metal oxide gas sensors. This initiative not only fills a gap in existing research, which has largely relied on lab-grade equipment for chemical analysis, but also democratizes access to data for building gas/odor-based emotion recognition models applicable in everyday settings.

Our paper offers the following contributions: 
(1) The \emph{AAQ} dataset: the first open-access collection of personal chemical emission data from gas sensors (wearable and desktop), capturing participant responses to 12 videos across four emotions (Amusement, Relaxation, Sadness, Disgust) during a $\sim$47-minute controlled study.
(2) A simple, off-the-shelf design for a wearable gas sensing headset to roughly detect -- in real-time -- a variety of gases, besides \textit{carbon monoxide} (CO), \textit{nitrogen dioxide} (NO$_2$), \textit{ethyl alcohol} (C$_2$H$_5$CH), \textit{Volatile Organic Compounds} (VOC), etc. In addition, we provide qualitative results that point to the design's usability and robustness for sensing. 
(3) A preliminary analysis of automatic affect detection based on the data, which did not demonstrate full feasibility but held promising results pointing to a need for further investigation.
(4) An initial investigation of participant's perceptions on air quality monitoring, potential risks involving privacy and security, as well as ethical considerations when designing AAQ systems. More specifically, we also provide and analyze 3 hours and 50 minutes of transcribed exit interviews on these topics.

Through a review of related work, design of a simple gas sensing wearable, detailed dataset construction, and preliminary analysis of findings, this paper lays the groundwork for further investigation into using gas sensors as a potential privacy-preserving, comfortable, and effective tool for emotion detection. By bridging the gap between breath \& body emission (emotional chemosignal) analysis and affective computing, the AAQ dataset aims to pave the way for advancements in alternative, unobtrusive emotion recognition technologies.
\section{Related Work}
To contextualize the contribution of this work, we provide an integrated overview of human emotional chemosignals, gas sensing technologies, and existing gas and air quality datasets.

\subsection{Emotional Chemosignaling in Humans}
Chemical signals, or chemosignals, significantly influence behavioral responses within species. The communicative function of chemosignals in humans, historically a topic of debate \cite{wysocki_facts_2004, doty_great_2010}, has gained clarity over recent decades. Emerging research underscores chemosignals' critical role in intra- and interspecific interactions \cite{calvi_scent_2020}, particularly emphasizing the conveyance of emotional states through bodily emissions \cite{de_groot_chemosignals_2012}. Central to this discourse is the notion that human emotions materialize physically, notably altering body odor and breath composition. Far from mere physiological byproducts, these changes act as nuanced communicative cues, shaping social perceptions and interactions. Intriguingly, common social behaviors, such as handshaking, may facilitate the unconscious sampling of another's chemosignals, evidenced by increased hand sniffing post-handshake \cite{frumin_social_2015}.

Historically, research into emotional chemosignaling has predominantly explored the transmission of ``negative emotions,'' such as fear, stress, or anxiety \cite{de_groot_human_2017}. For instance, fear-induced sweat can bias others toward perceiving fear in ambiguous facial expressions \cite{zhou_fear-related_2009}, a phenomenon observed regardless of the fear-associated chemosignal concentration and not observed with sweat induced by physical exercise \cite{de_groot_titrating_2021}. Similarly, male recipients exposed to women's tears reported diminished sexual arousal, an effect absent when exposed to odorless, negative-emotion tears \cite{gelstein_human_2011}.

Conversely, further research has recorded analogous outcomes with ``positive emotions,'' such as happiness or increased sexual arousal \cite{iversen_enhanced_2015, zhou_entangled_2011, zhou_reduced_2011, de_groot_sniff_2015}, indicating a broad spectrum of emotions communicable via human chemosignals.

While the primary focus has been on body odors (i.e., sweat), breath composition may also contain emotional chemosignals, as composition changes after stressful or relaxing stimuli \cite{camarinha-matos_real_2020} and even repeatably throughout a film when analyzing an audience's aggregated breath composition \cite{williams_crowd-based_2016}.

Given the complex chemical composition of these emissions, behavioral scientists have called for machine learning techniques to correlate chemical structures with psychological effects  \cite{de_groot_more_2020}. This work aims to provide the affective computing field with the data to explore the link between bodily emission and emotional state to further the investigation of data analysis and machine learning approaches for affective air quality.

\subsection{Gas Sensing Technologies}
Metal-oxide (MOX) gas sensors, or chemiresistors, are central to smell-sensing applications due to their cost-effectiveness and compactness. They detect changes in electrical resistance when a metal oxide surface absorbs oxygen in the presence of gases \cite{wang_metal_2010}, making them a practical choice for integrating into interactive systems. By adjusting the surface coating, thickness, or shape, MOX sensors can be fine-tuned to detect specific gases, though they generally react to a wide range of gases non-selectively.

Gas sensors have been employed to identify everyday activities \cite{kobayashi_context_2011} such as eating, restroom use, and smoking, and have monitored cooking stages and household scents, including cooking odors and candle lighting.\cite{hirano_detecting_2013, khaloo_nose_2019, kratz_whats_2022}.
Projects like \emph{Bin-ary} have used gas sensors for environmental applications, such as masking organic waste malodors \cite{amores_bin-ary_2015}, and \emph{AirSense} monitored indoor air quality \cite{fang_airsense_2016}, showcasing the versatility of gas sensing in enhancing environmental and quality of life.

One of the earliest HCI research example of gas sensing for interaction is \emph{Back to Mouth}, an interactive game using breath odors to repulse monsters \cite{iwamoto_back_2009}. Recent initiatives propose electronic noses for assessing body odor intensity \cite{eamsa-ard_development_2018}and incorporate gas sensors into wearables for health, well-being, and environmental monitoring, like \emph{VOCNEA} for sleep apnea detection \cite{roddiger_vocnea_2018} and \emph{AirSpec} smart glasses to track the impact of air quality on comfort \cite{chwalek_airspec_2023}. Multiple projects have also integrated gas sensors to measure personal exposure to air pollution both indoor and outdoor \cite{dam_wearable_2017,maag_w-air_2018,geczy_wearable_2020,zhong_hilo-wear_2020,palomeque-mangut_wearable_2022}.

While existing research in gas sensing within HCI has primarily concentrated on detecting odors or monitoring our environment, we sought to produce a dataset exploring the connection between a user's breath or bodily emission composition and their emotional state.

\subsection{Breath and body odor analysis}
Detecting emotions via body odor or a breath’s chemical composition is not new. Using mass spectrometry, researchers applied atmospheric chemistry concepts to sensitively measure the changes in bodily emanations from humans in isolation or groups \cite{roberts_decoding_2020}. Others have explored the use of quartz crystal resonators with plasma-polymer films tuned for specific chemical compounds to differentiate between a breath from a stressed or relaxed participant with an accuracy of 70\% \cite{takahashi_feasibility_2008, takahashi_remarks_2009}. However, the relaxed condition had participants drinking tea, which may have impacted the breath composition for this condition. Still, others measured breath composition changes after long stressful or relaxing stimuli using a gas chromatographic-ion mobility spectrometer \cite{camarinha-matos_real_2020}.

In contrast to the measurement of individuals, others have proposed analyzing the collective breath of crowds to assess group behavior, activity, or emotions (``crowd breath'') \cite{williams_crowd-based_2016}, building off their prior work showing repeatable changes in measured VOC in audience breath in a cinema while watching movies \cite{wicker_cinema_2015,williams_cinema_2016}. They used a mass spectrometer to detect and identify precise volatile compound concentrations. Most recently, a study demonstrated that a crowd’s collective experience directly correlates to the biomarkers in the crowd’s breath \cite{bensemann_what_2023}, suggesting visual and auditory experiences have predictable effects on human volatile compounds.

However, the use of mass spectrometers for such analyses presents challenges, including their size, cost, and the need for specialized operation. For crowd breath studies, these devices require a building-wide ventilation system to collect samples, introducing a five-minute delay for volatile compounds to reach the spectrometer.

Unlike prior work, we present a dataset of real-time emotional responses to video clips using an individual's breathing via cheap, off-the-shelf MOX gas sensors.

\subsection{Gas Sensing Datasets}
Research in environmental monitoring and electronic nose applications has generated datasets that, despite often utilizing similar technology—arrays of low-cost commercial metal oxide (MOX) gas sensors—differ significantly in their focus and methodology. Environmental monitoring datasets typically aim to quantify air pollutants like\emph{carbon monoxide} (CO), \emph{carbon dioxide} (CO$_2$), \emph{sulfur dioxide} (SO$_2$), \emph{nitrogen dioxide} (NO$_2$), and \emph{ozone} (O), focusing on air quality assessments \cite{wei_one-year_2023,betancourt_aq-bench_2021,barcelo-ordinas_h2020_2021,ferrer-cid_raw_2022}. Conversely, electronic nose datasets, prioritizing a broader array of MOX sensors, target creating ``digital fingerprints'' of gas mixtures for applications in food safety, health, and gas discrimination, offering insights into the sensor behavior across various stages of food spoilage, quality, or adulteration \cite{rodriguez_gamboa_electronic_2019-1, wijaya_electronic_2018, erlangga_electronic_2021, wijaya_electronic_2022, sarno_electronic_2020-1}, as well as identifying chemical compounds and studying sensor drift \cite{qu_open-set_2022, fonollosa_chemical_2015}. A unique dataset also explores electronic nose responses to breath samples from healthy individuals and those with Chronic Obstructive Pulmonary Disease (COPD) \cite{duran_acevedo_electronic_2021-1}. However, there appears to be no publicly available dataset yet that captures gas sensor data correlated with diverse emotional experiences.
However, as far as we have found, there is currently no publicly accessible dataset collecting gas sensor data during emotional experiences.
\begin{figure*}[t]
\centerline{\includegraphics[width=1\linewidth]{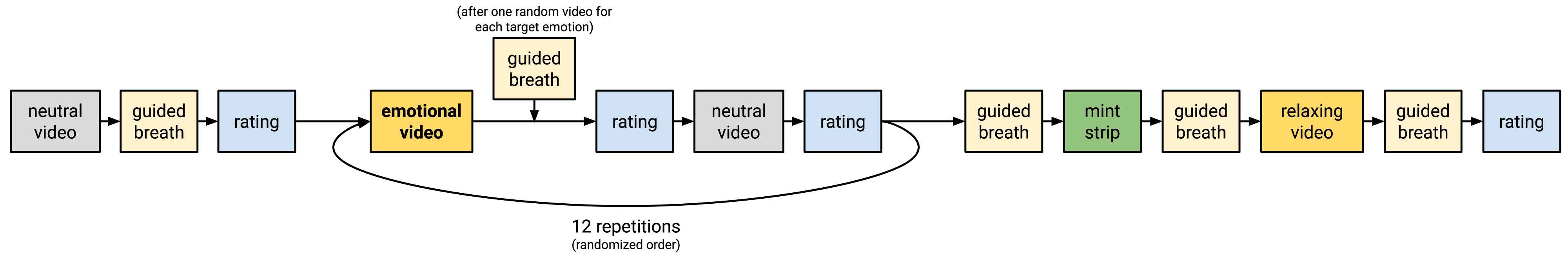}}
\caption{The core of the experiment consisted of twelve iterations, each including: (1) viewing an emotion-eliciting video clip, (2) completing a self-assessment, (3) watching a 30-second neutral washout clip (to minimize carryover effects), and (4) completing another self-assessment. In addition, at the end of certain emotional clips and first neutral video, participants were guided to hold their breath and exhale for three seconds. Before the final clip, half of the participants took a Listerine Cool Mint breath strip and let it dissolve on their tongue. Then, all participants watched a relaxing clip followed by a final self-assessment.}
\label{fig:StudyFlow}
\end{figure*}

\section{The Affective Air Quality Dataset}

The \emph{Affective Air Quality} (AAQ) Dataset consists of emotional ratings and 18 hours \& 11 minutes of gas data collected from 23 participants while watching 12 movie clips that targeted five emotional responses (amusement, disgust, relaxation, sadness, and neutral), with a total of 3 clips per valance-arousal quadrant. The data was collected using a wearable device we developed using an off-the-shelf air quality monitor with four gas sensors (described in Section \ref{gasensor}). We also developed an annotation platform to present the movie clips in a randomized order, which allowed the participant to rate their emotional response after each clip using 9-item Likert scales for valence (from very unpleasant to very pleasant), arousal (low to high energy), and familiarity with the content. Section \ref{emoclips} summarizes the stimuli to elicit targeted emotional responses. In addition to the gas sensor data and emotional ratings, we provide transcripts from all exit interviews ($\approx$4h of data), as well as (1) the timing and nature of their last consumed beverage and meal, (2) sleep quality via the Richards-Campbell Sleep Questionnaire, (3) personality traits through the short Big Five personality test, (4) current and general anxiety levels measured by the short State-Trait Anxiety Inventory (STAI), (5) mood states assessed by the short form of the International Positive and Negative Affect Schedule (I-PANAS-SF) and (6) emotional regulation skills via the ERQ Questionnaire.

Detailed information about participant's demographics, experimental setup, and data collection can be found in the following sections and appendix. All data collected has been made anonymous to prevent potential privacy breaches. The study was approved by the local Institutional Review Board and conducted in a controlled office room from August 15, 2023, to September 12, 2023.

\subsection{Participants and Data Exclusion}
We recruited 34 healthy participants aged 18-64 through email and advertising, but due to recording software issues, data from 11 sessions were lost, though their exit interviews were preserved. This left us with data from 22 complete sessions and 1 partial session due to the sensing apparatus accidentally getting disconnected. Participant ages were: 18--24 ($n=7$), 25--34 ($n=14$), 45--54 ($n=1$), and 55--64 ($n=1$), with 16 identifying as women, 6 as men, and 1 as non-binary/gender diverse.

The exclusion criteria included a known history of smoking or lung and respiratory problems. Participants were required to have at least least 5 hours of sleep prior, no alcohol for 24 hours, no scented products, and no eating or drinking 2 hours before the study. The study was conducted in a controlled office room,  and all participants provided written informed consent and were compensated with a 70 USD gift card.

At the start of the study, participants completed an intake form and questionnaires (RCSQ, BFI, STAI, I-PANAS-SF) to document recent food or drink intake that could impact their bodily emissions and establish an emotional baseline.

\subsection{Experimental Procedure}
To ensure the integrity of our data, we scheduled adequate intervals between participants. This strategy allowed the building's ventilation system sufficient time to return the gas sensor readings to a stable baseline, ensuring no residual influences from prior participants lingered within the space. The study setup was initiated 30 minutes before each session to allow the air quality monitors to stabilize and establish a baseline reading of the room air composition. Upon arrival and after providing written consent, participants completed an intake form and were then led to wear the air quality headset (as shown in Figure \ref{fig:prototype}). 

We summarize the procedure in Figure \ref{fig:StudyFlow}. Participants were familiarized with the interface through a neutral video, during which they could adjust audio levels to their comfort. This introduction also covered the concepts of valence, arousal, and familiarity to prepare them for the self-assessments. Participants then watched 12 emotion-eliciting video clips (in a fully randomized order to mitigate potential ordering effects) and rated their valence, arousal, and familiarity after each. Neutral washout clips were randomly selected from two options to maintain participant engagement. An additional (and final) relaxing video aimed to help mitigate any residual negative emotions induced by previous stimuli, ensuring participants ended the experiment in a positive or neutral emotional state. Before this final clip, half of the participants received a mint-flavored breath strip in order to potentially examine if breath alteration could negatively impact emotion recognition using these sensors. (This portion was inspired by the peppermint benchmarking protocol for  GC-MS-based breath analysis \cite{henderson_benchmarking_2020,wilkinson_peppermint_2021}.) After this portion, the researcher engaged the participant in a semi-structured interview. The interview focused on feedback on the equipment, air quality monitoring experiences, and privacy concerns related to air quality data collection.

The total study duration was about 1 hour and 15 minutes.

\subsubsection{Hardware}\label{gasensor}
Chemical signals were captured using the Grove Multichannel Gas Sensor (v2), which includes four MOX gas sensors each optimized for different gases but generally non-selective towards oxidizing gases. These sensors are specifically responsive to \emph{nitrogen dioxide} (Winsen GM-102B), \emph{ethyl alcohol} (Winsen GM-302B), \emph{volatile organic compounds} (Winsen GM-502B), and \emph{carbon monoxide} (Winsen GM-702B), yet can detect a similar range of gases. With \emph{ethanol}, \emph{2-propranol} and \emph{1-propanol} identified as stress-sensitive compounds \cite{camarinha-matos_real_2020}, MOX sensors hold potential for stress detection.

\begin{figure}[ht]
\centerline{\includegraphics[width=0.7\columnwidth]{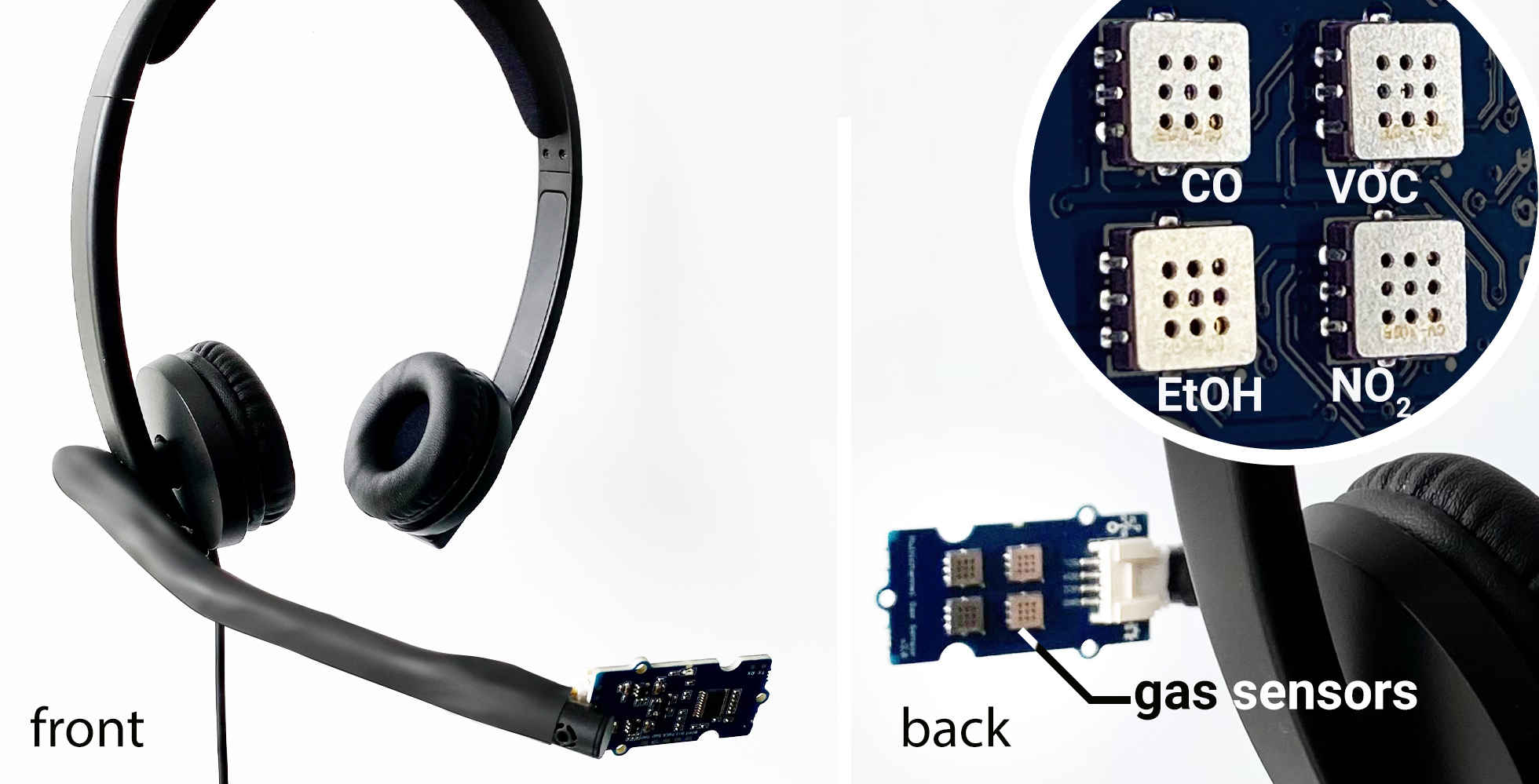}}
\caption{Prototype wearable gas sensor system using headphones, with sensors mounted on the microphone goose-neck near the nose and mouth. The sensors are optimized for response to \emph{nitrogen dioxide} (NO$_2$), \emph{ethyl alcohol} (EtOH), volatile organic compounds (VOC), and \emph{carbon monoxide} (CO).}
\label{fig:prototype}
\end{figure}

To ensure thorough coverage of the testing area, we placed gas sensors in two strategic locations: directly in front of the participant's nose and mouth, and on the study desk about $\SI{0.75}{\meter}$ from the participant. The sensor near the nose and mouth was attached to a microphone goose-neck on standard headphones (detailed in Figure \ref{fig:prototype}). Participants were guided on aligning the microphone for optimal capture of their breath, though most managed this without help.

\subsubsection{Software}
We utilized Qualtrics for intake form questionnaires and employed two scripts for data collection during emotional video clips. A Python script managed the recording and timestamping from both air quality monitors at a sampling rate of $\SI{91}{\hertz}$, except for P11’s session, recorded at $\SI{58}{\hertz}$ due to software differences. Additionally, we developed an annotation platform with PsychoPy to oversee the experimental sequence and let participants' annotate their emotional responses after each video.

\subsection{Emotional Movie Clips}\label{emoclips}
\begin{figure}[t]
\centerline{\includegraphics[width=0.8\columnwidth]{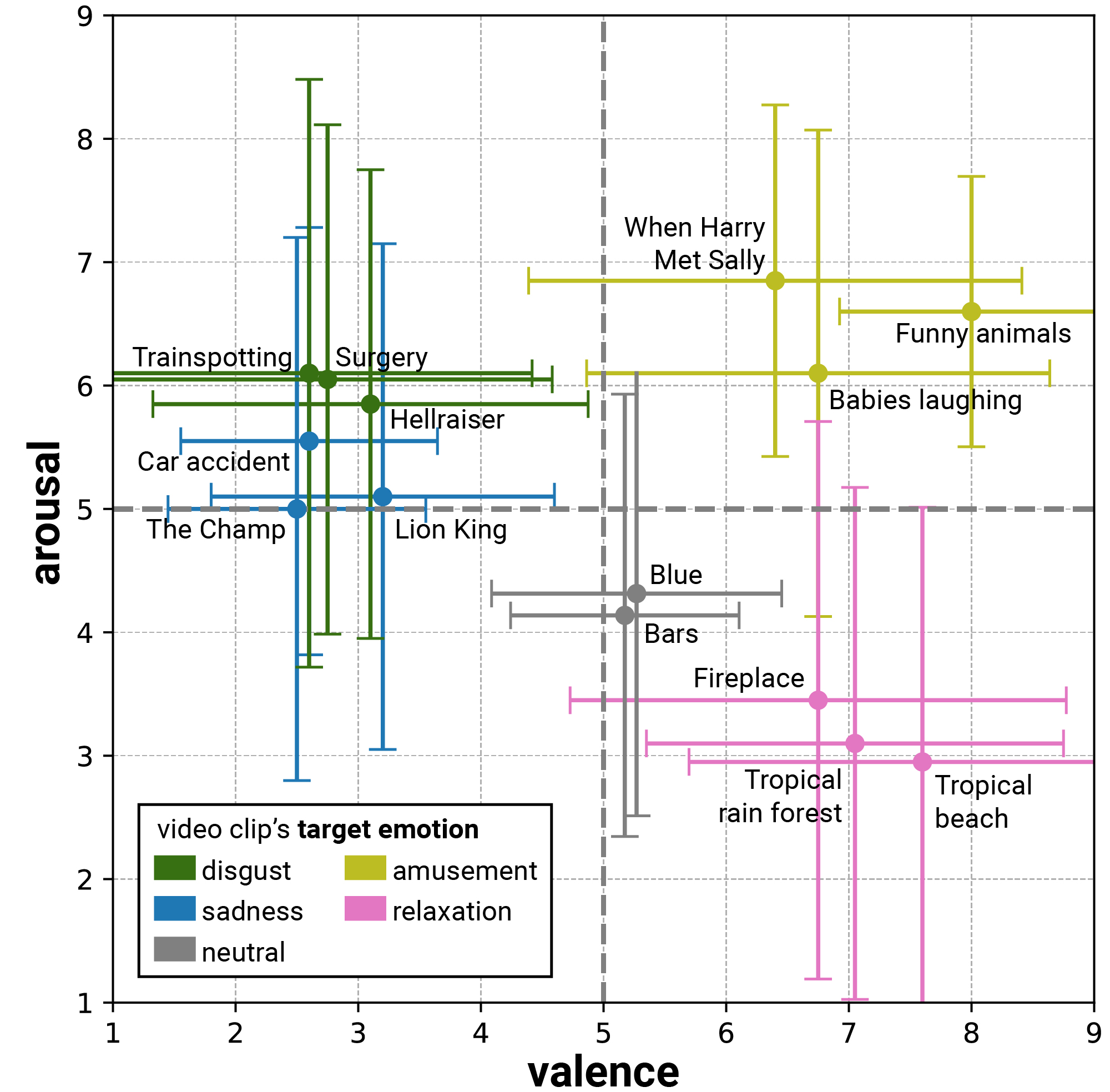}}
\caption{Average valence and arousal (and their standard deviations) for each presented emotion-eliciting video. Reported valence and arousal largely aligned with the targeted emotion, though Sadness-eliciting videos  drifted towards a higher arousal state and Neutral videos drifted to negative valence from repeated exposure.}
\label{fig:VAS}
\end{figure}

In order to elicit targeted emotional responses, we utilized short video clips sourced from databases with established reliability and validity \cite{rottenberg_emotion_2007, schaefer_assessing_2010, saganowski_emognition_2022}. Recognizing the prior literature's emphasis on negative valence emotions such as Disgust, Fear, and Sadness, we incorporated clips known to induce positive valence such as Relaxation \cite{benz_nature-based_2022} and Amusement as shown in Figure \ref{fig:VAS}. For Amusement we sourced high quality funny animal video clips from YouTube based on their similarity to what others have used in the past \cite{soleymani_multimodal_2012}, prioritizing clips with over six million views by August 2023, and search terms including ``funny video,'' ``babies laughing,'' and ``hilarious videos.'' \looseness=-1

To balance stimuli across the valence-arousal space, we selected enough footage in each quadrant to reach 399 seconds of stimuli across three videos. To balance the Disgust category, we appended additional surgery footage to the surgery video originally included in \cite{rottenberg_emotion_2007}. Each quadrant of the valence-arousal space was limited to three video clips, standardizing the number of washout neutral videos across the experiment. All selected videos met a minimum resolution criterion of {1280$\times$720}\,\si{\pixel}. Ultimately, 12 emotion-eliciting clips were chosen, with lengths ranging from 59 to 232 seconds long ($M = \SI{133}{\second}$, $SD = \SI{45.6}{\second}$). Detailed information about the video clip sources and characteristics is provided in our repository. \looseness=-1

\section{Data Records and Contents}
The AAQ dataset is accessible on GitHub.\footnote{\url{https://github.com/microsoft/aaq}} The PsychoPy experimental setup code is also accessible as a Supplementary Material. Due to the lack of standard preprocessing pipelines for continuous gas sensing of human breath, we include both raw data and a preliminary quantitative analysis with details on preprocessing and feature extraction steps. The dataset contents are summarized in Appendix Table \ref{tab:files}.
\begin{figure}
\centerline{\includegraphics[width=1\linewidth]{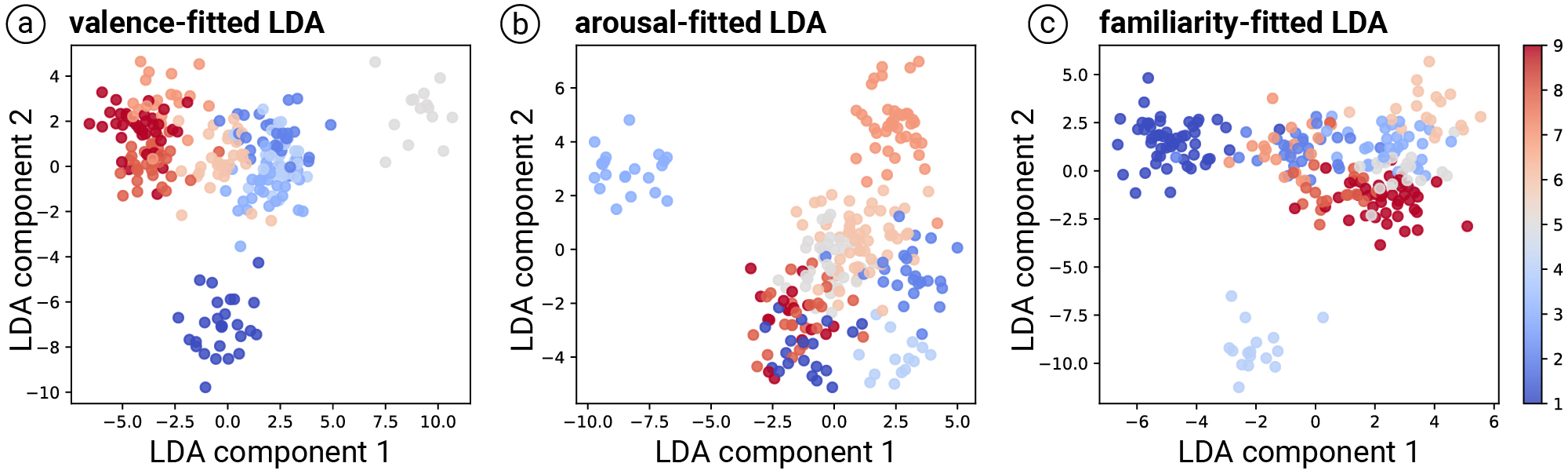}}
\caption{2D Linear Discriminant Analysis (LDA) projections based on features extracted from sensor data, separately fitted for (a) valence, (b) arousal, and (c) familiarity. All three plots demonstrate clear clustering for scores.}
\label{fig:LDAs}
\end{figure}

\section{Gas Sensor Analysis}

To set a baseline for the community, we conducted a preliminary investigation into the possibility of automatic emotion detection with the AAQ dataset.

\subsection{Preprocessing and Feature Extraction}
To prepare our gas sensor data for feature extraction, we implemented several pre-processing steps. We used a fourth-order band-pass Butterworth filter with a $\SI{30}{\hertz}$ cutoff to eliminate high and low-frequency noise, followed by an Exponential Weighted Moving Average (EWMA) with $\alpha = 0.08$. Finally, min-max normalization was applied to standardize the data range across channels, per video clip.

Given the novelty of the set-up, it is unclear which features would be most relevant to detecting emotion from gas sensor signals. While we do not track participants' respiration, we propose that the averaged signal across the four headset gas sensors could be used to estimate respiration cycles, inspired by VOCNEA \cite{roddiger_vocnea_2018}. Alternatively, gas sensor signals during a single exhalation or full respiration cycle could also be used.

As a first step, we included the gas sensor signals from both the headset and desktop AQ monitors during emotion-eliciting clips. Features extracted included statistical measures (mean, standard deviation, area under curve, skewness, kurtosis), conductance changes, relative abundance (separated by device), mean \& standard deviation of each DWT coefficient, and R\'enyi entropy for each gas sensor. Additionally, we sensor response metrics (peak-to-peak time, trough-to-peak signal, and minimum and maximum slope) for both the unnormalized and normalized gas sensor signals, generating in total a $\SI{248}{D}$ feature vector. Inspired by \cite{li_recognizing_2020}, we share the Linear Discriminant Analysis (LDA) plots to present the separability of arousal, valence, and familiarity scores (see Figure \ref{fig:LDAs}).

For emotion prediction, we employed SVM (with RBF kernel), RF (100 trees), and XGBoost models as well as an LSTM network, using a 5-fold cross-validation approach. After initial testing with LDA, we opted to not conduct dimensionality reduction for the SVM, RF, and XGBOOST model. We treated the problem as a regression task, for which the models needed to predict the valence or arousal of a sequence (we trained separate models for each). Our LSTM model, designed to address the data's temporal aspect, included two 100-unit layers, processed the last 30 seconds of sensor signals, and featured Dropout layers to reduce overfitting. The network used the GlorotUniform initializer and LeakyReLU activation, concluding with a Dense layer for valence or arousal predictions. Training involved the Adam optimizer and mean squared error loss, with early stopping to curtail overfitting after 50 epochs or 10 epochs without loss improvement. For the cross-validation, all samples were randomly divided into five folds, and in each validation, four folds are used for training and the remaining fold for testing. (Note, the data from the same participant may be divided across the training and testing sets, which is a current limitation of our preliminary analysis.)

\subsection{Results}
For evaluating our models, we prioritized Mean Absolute Error (MAE) and Root Mean Squared Error (RMSE) as our metrics, given their direct interpretability in the context of numerical valence and arousal prediction. MAE provides a clear measure of average prediction accuracy in the same units as the target variable, making it particularly relevant for applications in affective computing where precision in emotional state prediction is paramount. The RMSE complements the MAE metric by providing a relatively high weight to large errors, making it useful to evaluate a model when large errors are particularly undesirable. Finally, we included the R$^2$, which represents the proportion of the variance in the dependent variable that is predictable from the independent variable.

\begin{table*}[htb]
\caption{Summary of model performance metrics from a 5-fold cross validation.}
\begin{center}
\scriptsize
\begin{tabular}{@{}l|ccc|ccc|@{}}
\toprule
        & \multicolumn{3}{c|}{\textbf{Valence}}             & \multicolumn{3}{c|}{\textbf{Arousal}} \\ \midrule
Model   & MAE & RMSE & R$^2$ & MAE    & RMSE    & R$^2$    \\ \midrule
SVM     & $2.30\pm0.17$    & $2.64\pm0.19$     & $0.03\pm0.05$      & $1.85\pm0.27$       & $2.28\pm0.28$        & $-0.03\pm0.07$         \\
RF      & $1.93\pm0.15$    & $2.32\pm0.19$     & $0.25\pm0.08$      & $1.80\pm0.31$       & $2.17\pm0.30$        & $0.07\pm0.04$         \\
XGBOOST & $2.09\pm0.23$    & $2.57\pm0.23$     & $0.06\pm0.19$      & $1.82\pm0.28$       & $2.22\pm0.28$        & $0.02\pm0.16$         \\
LSTM    & $2.43\pm0.23$    & $2.72\pm0.18$     & $-0.05\pm0.06$      & $1.90\pm0.22$       & $2.33\pm0.28$        & $-0.07\pm0.13$         \\ \bottomrule
\end{tabular}
\label{tab:metrics}
\end{center}
\end{table*}

Table \ref{tab:metrics} shows the prediction results of the preliminary models trained. From the table, we can see that model predictions were typically within two units of the actual valence arousal levels on a scale from 1 to 9. The models' RMSE averaging around 2-3 across folds further supports their efficacy by indicating low dispersion of errors. Unfortunately, most models do not explain the full variance in valence or arousal (R$^2$), though the RF model does notably better.
\section{Interviews}
We collected approximately 3 hours and 50 minutes of transcribed interviews across all participants\footnote{For enhanced clarity in qualitative analysis, we have reassigned participant identifiers. Detailed information on these updated labels can be found in the Participant Logs section of our repository.} (including those whose quantitative sessions were lost). Using Burnard's method \cite{burnard_method_1991}, we identified key topics such as the privacy and sensitivity of AAQ data, focusing on social intimacy concerns of bodily scents and the perception of air quality data as non-private. We also explored the connection between emotion and health information, sensitivity towards air quality data, usability feedback, strategies in monitored environments, and the use of AQ-related technologies.

All participants found the headset comfortable or unobtrusive. P15 stated, ``It's comfortable. I like it,'' while P31 found the prototype familiar and comfortable, ``it's very familiar to the headset I usually wear.'' Three participants mentioned slight discomfort from the headset pressing against their glasses, several suggested improving the prototype by using wireless headphones that cover the ear, and P27 suggested integrating a longer microphone goose-neck to better adjust the sensors' position.

\subsection{Social Intimacy of Bodily Emissions}
In our exploration of user perceptions regarding monitoring bodily emissions and odors, a theme of intimacy emerged, intertwined with notions of social privacy and personal identity. Our findings suggest that while there is curiosity about the potential insights derived from monitoring bodily emissions, there remains a significant concern for the intimacy and social implications of such data collection.Our analysis reveals a key misunderstanding about air quality (AQ) monitoring technology capabilities. Participants often conflate the quantitative collection of data regarding bodily emissions with the qualitative recording of the \emph{actual scent}, fearing that such technology could enable judgments about the odor itself (and -- by extension -- themselves). As P24 intimates, the worry is not just about data collection but about the potential for this data to reflect on them through a lens of social judgment. This concern highlights a critical gap in understanding; current technology does not allow for the social assessment of odors in the way participants fear (poor AQ readings do not inherently mean bad smells), yet the anxiety about being subject to such scrutiny remains palpable.

\subsection{Privacy and Sensitivity of Air Quality Data}
Despite concerns about the social implications of monitoring bodily emissions, a majority participants ($n=16$) did not perceive AQ data related to their bodies as private or sensitive. This sentiment was echoed across responses, suggesting a comfort with the collection and use of AQ data for purposes such as public health and personal wellness.

Many participants said they do not consider AQ data related to their body as personally identifiable or sensitive. For instance, participants likened sharing AQ data to sharing non-sensitive, publicly observable information, such as body odor or breath quality, or to everyday experiences where personal scents are naturally detected by others in close proximity. These comparisons imply a belief that AQ data does not constitute a serious privacy concern.

Several participants explicitly dismissed concerns over the privacy of AQ data. For example, P15 found value in accessing any internal information for health purposes (``Not really... it's always good to have access''), while P9 expressed no particular concern, and P16 likened it to ``just the air.'' P22 did not consider AQ data as personally identifiable or a privacy issue, assuming responsible data management.

A subset of participants differentiated between types of data, deeming some as more privacy-sensitive than others. AQ data was often seen as less sensitive than personal data such as health records. This perspective was grounded in the belief that AQ, a component of the surrounding environment, inherently carries less personal identification potential.


These reflections collectively illustrate a predominant sentiment among participants that AQ data, particularly when related to bodily emissions, does not significantly intrude upon personal privacy. This consensus suggests a potential broader acceptance of the collection and use of such data.

However, privacy concerns were conditional for several participants ($n=11$), depending on the context of data use and the extent of anonymization. There was a recognition of the utility of AQ data for public health or environmental monitoring, with several participants indicating acceptance of collective data use for broader societal benefits. This acceptance was, however, contingent on assurances of anonymity and appropriate use, highlighting a nuanced understanding of privacy concerns where individual consent is balanced against the potential for communal benefit.

15 participants evaluated the perceived sensitivity of not only AQ data but also emotional and health data derived from the AQ data. Through our coding process, we assigned these perceptions into four broad categories: Not Sensitive/Private (0), Slightly Sensitive/Private (1), Conditionally Sensitive/Private (2), and Sensitive/Private (3). Our analysis revealed a positive correlation between these sensitivities, indicating that concerns about privacy tend to increase together. Specifically, we found a moderate positive correlation (0.37) between AQ and Emotional sensitivity, and a stronger correlation (0.54) with Health sensitivity.

Among participants who considered AQ data not sensitive or private ($n=8$), about half also viewed derived Emotion information as not sensitive. Roughly a third felt the same about Health information derived from AQ.

These results suggest a trend: individuals who already regard AQ data as sensitive and private are more inclined to view Emotional and Health data through a similar lens of sensitivity and privacy, with an especially pronounced correlation observed in the realm of Health data.

\subsection{Strategies for Privacy Protection}
In discussions on privacy in AQ-monitored environments, seventeen participants devised strategies to safeguard their privacy, reflecting a range of concerns and pragmatism. Three participants considered using perfumes, breath mints, or hygiene products to mask bodily emissions, with one viewing it as introducing "artifacts" into the data. Four others thought about altering their breathing or using masks to affect data collection. Some even considered tampering with the equipment directly, such as covering or disabling sensors.

However, responses to these privacy measures varied. Two participants deliberated the effort versus the privacy benefit, showing a conditional approach. Meanwhile, a few displayed indifference towards privacy in these scenarios.

\subsection{Personal Use of AQ-related Devices}
After consolidating responses to account for variations in terminology, we identified the following as the most frequently cited AQ-related systems used by participants: \emph{smoke detectors} (30), \emph{carbon monoxide detectors} (17), \emph{humidifiers or dehumidifiers} (11), and \emph{air purifiers} (9). Air purifiers and humidifiers, often situated in bedrooms, are regularly interacted with directly. In contrast, smoke alarms and carbon monoxide alarms are more passive fixtures, spanning many rooms and generally only drawing attention when triggered.
\section{Discussion}
While our models' R$^2$ were largely small (except for the RF model's R$^2$ for valence), the MAE and RMSE's coarse accuracy suggest an interesting potential for the modality's use in affective computing applications. The models may be suitable to roughly estimate levels in valence and arousal (i.e., low, neutral, and high), but not fine-grained prediction. A future direction could involve training an LSTM network on the features we used for the RF models, but with a sliding window capturing 2-3 breathing cycles. 

The Linear Discriminant Analysis (LDA) plots (Figure \ref{fig:LDAs}) suggest that there is space for further enhancing separation between emotional states with improved feature engineering.  Notably, LDA showed improved separation when combining data from both headset and desktop sensors and including trough/peak features, suggesting these could be fruitful areas for further research. Ultimately, our preliminary analysis leaves room for many improvements.

Our qualitative analysis revealed two interesting lines of thought. First, while participants may find bodily emissions intimate or socially concerning, most participants paradoxically did not view AQ data as sensitive. Our interviews also identified a network of AQ sensors in domestic and professional environments, potentially laying the groundwork for future affective computing research utilizing bodily emissions. Despite their lack of proximity to the body compared to devices in our dataset, their closeness during prolonged activities like sleep could allow for ongoing emission monitoring, complementing wearable technologies.

However, this study has limitations. Our participant pool was predominantly young women, indicating a need for greater diversity. We lacked specific breath monitoring and environmental data like temperature and humidity, though we propose a method for estimating breathing patterns for future exploration. The emotional response overlap in our videos, particularly between disgust and sadness, underscores the need for clearer differentiation in stimuli. Our cross-validation method, which might include data from the same participants in both training and testing sets, raises concerns about the models learning individual characteristics rather than generalizable features. Additionally, the timing of our exit interviews may have influenced participants' responses to privacy concerns.

Lastly, we introduced a relaxation video with breath mint masking but have yet to assess its impact on model performance in detecting valence or arousal. This aspect, along with the aforementioned limitations and observations, offers rich avenues for further investigation by the research community.

\section{Conclusion}
In this work, we introduced the \emph{Affective Air Quality} dataset to explore affect recognition through the novel lens of emotional chemosignaling using off-the-shelf air quality monitors. This dataset comprises 4-channel gas sensor data collected from two positions (head-mounted and desktop) across 23 individuals experiencing varying emotions elicited by video clips. Additionally, it includes valence, arousal, and familiarity assessments for each clip, and is complemented by exit interview transcripts that shed light on participants' views on air quality monitoring and its implications for privacy.

Our work not only introduces a novel dataset but also lays the groundwork for future explorations in affect recognition utilizing this non-visual, non-contact method. The findings suggest that gas sensor data could be used to correlate changes in breath or body odor with emotional states, an area of research currently needing sophisticated and costly laboratory equipment like mass spectrometers, or methods requiring structured breathing protocols. We encourage the research community to engage with our dataset to develop affect recognition models and engineer even better features for extraction, opening new avenues for understanding and interpreting human emotions through the lens of human chemosignaling.
\looseness=-1

\section*{Ethical Impact Statement}
Emotion recognition technologies inherently carry a significant risk of misuse, given the sensitivity of emotional information. Our protocol incorporated semi-structured exit interviews to gauge participant perceptions regarding the privacy and sensitivity of air quality data, including how it relates to the inference of emotional states or personal health information. As evidenced by our qualitative findings, participants paradoxically have concerns about their body odor or breath being monitored (from fear of social judgment about the smells' hedonic qualities), yet are generally not concerned about air quality data's sensitivity or privacy unless it specifically discloses extracting emotional or personal health information. This paradox underscores a critical ethical consideration: the need for clear communication and stringent privacy safeguards when developing and deploying technologies that intersect closely with emotion, including when using non-contact and non-visual methods like AAQ.

Given the sensitive video clips often included in emotion-eliciting video studies (e.g., disgust- or anger-eliciting videos), we took much care to make the protocol non-distressing. Firstly, we ensured participants knew they could stop or exit the videos at any time, though no one opted to. Secondly, to protect participant well-being, we deliberately omitted videos that induce anger through depictions of racist or sexist violence, focusing instead on disgust-eliciting clips for low-valence, high-arousal emotions. We balanced the emotional range by including relaxing videos for high-valence, low-arousal emotions. Finally, we presented a relaxing video at the end of the protocol to help mitigate any residual negative emotions induced by previous stimuli, ensuring participants ended the experiment in a positive or neutral emotional state. We suggest future researchers employing emotion-eliciting videos to carefully select their content, as it may profoundly distress participants, especially those with personal connections to the material presented.

\bibliographystyle{IEEEtran}
\bibliography{AAQ_bib} 

\begin{thebibliography}{10}
\providecommand{\url}[1]{#1}
\csname url@samestyle\endcsname
\providecommand{\newblock}{\relax}
\providecommand{\bibinfo}[2]{#2}
\providecommand{\BIBentrySTDinterwordspacing}{\spaceskip=0pt\relax}
\providecommand{\BIBentryALTinterwordstretchfactor}{4}
\providecommand{\BIBentryALTinterwordspacing}{\spaceskip=\fontdimen2\font plus
\BIBentryALTinterwordstretchfactor\fontdimen3\font minus \fontdimen4\font\relax}
\providecommand{\BIBforeignlanguage}[2]{{%
\expandafter\ifx\csname l@#1\endcsname\relax
\typeout{** WARNING: IEEEtran.bst: No hyphenation pattern has been}%
\typeout{** loaded for the language `#1'. Using the pattern for}%
\typeout{** the default language instead.}%
\else
\language=\csname l@#1\endcsname
\fi
#2}}
\providecommand{\BIBdecl}{\relax}
\BIBdecl

\bibitem{camarinha-matos_real_2020}
P.~Santos, P.~Roth, J.~M. Fernandes, V.~Fetter, and V.~Vassilenko, ``Real {{Time Mental Stress Detection Through Breath Analysis}},'' in \emph{Technological {{Innovation}} for {{Life Improvement}}}, L.~M. {Camarinha-Matos}, N.~Farhadi, F.~Lopes, and H.~Pereira, Eds.\hskip 1em plus 0.5em minus 0.4em\relax Cham: Springer International Publishing, 2020, vol. 577, pp. 403--410.

\bibitem{de_groot_human_2017}
J.~H.~B. De~Groot and M.~A.~M. Smeets, ``Human {{Fear Chemosignaling}}: {{Evidence}} from a {{Meta-Analysis}},'' \emph{Chemical Senses}, vol.~42, no.~8, pp. 663--673, Oct. 2017.

\bibitem{de_groot_sniff_2015}
J.~H.~B. De~Groot, M.~A.~M. Smeets, M.~J. Rowson, P.~J. Bulsing, C.~G. Blonk, J.~E. Wilkinson, and G.~R. Semin, ``A {{Sniff}} of {{Happiness}},'' \emph{Psychological Science}, vol.~26, no.~6, pp. 684--700, Jun. 2015.

\bibitem{calvi_scent_2020}
E.~Calvi, U.~Quassolo, M.~Massaia, A.~Scandurra, B.~D'Aniello, and P.~D'Amelio, ``The scent of emotions: {{A}} systematic review of human intra- and interspecific chemical communication of emotions,'' \emph{Brain and Behavior}, vol.~10, no.~5, p. e01585, May 2020.

\bibitem{alharbi_mask_2019}
R.~Alharbi, M.~Tolba, L.~C. Petito, J.~Hester, and N.~Alshurafa, ``To {{Mask}} or {{Not}} to {{Mask}}?: {{Balancing Privacy}} with {{Visual Confirmation Utility}} in {{Activity-Oriented Wearable Cameras}},'' \emph{Proceedings of the ACM on Interactive, Mobile, Wearable and Ubiquitous Technologies}, vol.~3, no.~3, pp. 1--29, Sep. 2019.

\bibitem{wysocki_facts_2004}
C.~J. Wysocki and G.~Preti, ``Facts, fallacies, fears, and frustrations with human pheromones,'' \emph{The Anatomical Record Part A: Discoveries in Molecular, Cellular, and Evolutionary Biology}, vol. 281A, no.~1, pp. 1201--1211, Nov. 2004.

\bibitem{doty_great_2010}
R.~L. Doty, \emph{The Great Pheromone Myth}.\hskip 1em plus 0.5em minus 0.4em\relax Baltimore: Johns Hopkins Univ. Press, 2010.

\bibitem{de_groot_chemosignals_2012}
J.~H.~B. De~Groot, M.~A.~M. Smeets, A.~Kaldewaij, M.~J.~A. Duijndam, and G.~R. Semin, ``Chemosignals {{Communicate Human Emotions}},'' \emph{Psychological Science}, vol.~23, no.~11, pp. 1417--1424, Nov. 2012.

\bibitem{frumin_social_2015}
I.~Frumin, O.~Perl, Y.~{Endevelt-Shapira}, A.~Eisen, N.~Eshel, I.~Heller, M.~Shemesh, A.~Ravia, L.~Sela, A.~Arzi, and N.~Sobel, ``A social chemosignaling function for human handshaking,'' \emph{eLife}, vol.~4, p. e05154, Mar. 2015.

\bibitem{zhou_fear-related_2009}
W.~Zhou and D.~Chen, ``Fear-{{Related Chemosignals Modulate Recognition}} of {{Fear}} in {{Ambiguous Facial Expressions}},'' \emph{Psychological Science}, vol.~20, no.~2, pp. 177--183, Feb. 2009.

\bibitem{de_groot_titrating_2021}
J.~H.~B. De~Groot, P.~A. Kirk, and J.~A. Gottfried, ``Titrating the {{Smell}} of {{Fear}}: {{Initial Evidence}} for {{Dose-Invariant Behavioral}}, {{Physiological}}, and {{Neural Responses}},'' \emph{Psychological Science}, vol.~32, no.~4, pp. 558--572, Apr. 2021.

\bibitem{gelstein_human_2011}
S.~Gelstein, Y.~Yeshurun, L.~Rozenkrantz, S.~Shushan, I.~Frumin, Y.~Roth, and N.~Sobel, ``Human {{Tears Contain}} a {{Chemosignal}},'' \emph{Science}, vol. 331, no. 6014, pp. 226--230, Jan. 2011.

\bibitem{iversen_enhanced_2015}
K.~D. Iversen, M.~Ptito, P.~M{\o}ller, and R.~Kupers, ``Enhanced {{Chemosensory Detection}} of {{Negative Emotions}} in {{Congenital Blindness}},'' \emph{Neural Plasticity}, vol. 2015, pp. 1--7, 2015.

\bibitem{zhou_entangled_2011}
W.~Zhou and D.~Chen, ``Entangled chemosensory emotion and identity: {{Familiarity}} enhances detection of chemosensorily encoded emotion,'' \emph{Social Neuroscience}, vol.~6, no.~3, pp. 270--276, Jun. 2011.

\bibitem{zhou_reduced_2011}
W.~Zhou, P.~Hou, Y.~Zhou, and D.~Chen, ``Reduced recruitment of orbitofrontal cortex to human social chemosensory cues in social anxiety,'' \emph{NeuroImage}, vol.~55, no.~3, pp. 1401--1406, Apr. 2011.

\bibitem{williams_crowd-based_2016}
J.~Williams and J.~Pleil, ``Crowd-based breath analysis: Assessing behavior, activity, exposures, and emotional response of people in groups,'' \emph{Journal of Breath Research}, vol.~10, no.~3, p. 032001, Jun. 2016.

\bibitem{de_groot_more_2020}
J.~H.~B. De~Groot, I.~Croijmans, and M.~A.~M. Smeets, ``More {{Data}}, {{Please}}: {{Machine Learning}} to {{Advance}} the {{Multidisciplinary Science}} of {{Human Sociochemistry}},'' \emph{Frontiers in Psychology}, vol.~11, p. 581701, Oct. 2020.

\bibitem{wang_metal_2010}
C.~Wang, L.~Yin, L.~Zhang, D.~Xiang, and R.~Gao, ``Metal {{Oxide Gas Sensors}}: {{Sensitivity}} and {{Influencing Factors}},'' \emph{Sensors}, vol.~10, no.~3, pp. 2088--2106, Mar. 2010.

\bibitem{kobayashi_context_2011}
Y.~Kobayashi, T.~Terada, and M.~Tsukamoto, ``A {{Context Aware System Based}} on {{Scent}},'' in \emph{2011 15th {{Annual International Symposium}} on {{Wearable Computers}}}.\hskip 1em plus 0.5em minus 0.4em\relax San Francisco, CA, USA: IEEE, Jun. 2011, pp. 47--50.

\bibitem{hirano_detecting_2013}
S.~H. Hirano, J.~R. Brubaker, D.~J. Patterson, and G.~R. Hayes, ``Detecting cooking state with gas sensors during dry cooking,'' in \emph{Proceedings of the 2013 {{ACM}} International Joint Conference on {{Pervasive}} and Ubiquitous Computing}.\hskip 1em plus 0.5em minus 0.4em\relax Zurich Switzerland: ACM, Sep. 2013, pp. 411--414.

\bibitem{khaloo_nose_2019}
P.~Khaloo, B.~Oubre, J.~Yang, T.~Rahman, and S.~I. Lee, ``{{NOSE}}: {{A Novel Odor Sensing Engine}} for {{Ambient Monitoring}} of the {{Frying Cooking Method}} in {{Kitchen Environments}},'' \emph{Proceedings of the ACM on Interactive, Mobile, Wearable and Ubiquitous Technologies}, vol.~3, no.~2, pp. 1--25, Jun. 2019.

\bibitem{kratz_whats_2022}
S.~Kratz, A.~{Monroy-Hern{\'a}ndez}, and R.~Vaish, ``What's {{Cooking}}? {{Olfactory Sensing Using Off-the-Shelf Components}},'' in \emph{Adjunct {{Proceedings}} of the 35th {{Annual ACM Symposium}} on {{User Interface Software}} and {{Technology}}}.\hskip 1em plus 0.5em minus 0.4em\relax Bend OR USA: ACM, Oct. 2022, pp. 1--3.

\bibitem{amores_bin-ary_2015}
J.~Amores, P.~Maes, and J.~Paradiso, ``Bin-ary: Detecting the state of organic trash to prevent insalubrity,'' in \emph{Proceedings of the 2015 {{ACM International Joint Conference}} on {{Pervasive}} and {{Ubiquitous Computing}} and {{Proceedings}} of the 2015 {{ACM International Symposium}} on {{Wearable Computers}} - {{UbiComp}} '15}.\hskip 1em plus 0.5em minus 0.4em\relax Osaka, Japan: ACM Press, 2015, pp. 313--316.

\bibitem{fang_airsense_2016}
B.~Fang, Q.~Xu, T.~Park, and M.~Zhang, ``{{AirSense}}: An intelligent home-based sensing system for indoor air quality analytics,'' in \emph{Proceedings of the 2016 {{ACM International Joint Conference}} on {{Pervasive}} and {{Ubiquitous Computing}}}.\hskip 1em plus 0.5em minus 0.4em\relax Heidelberg Germany: ACM, Sep. 2016, pp. 109--119.

\bibitem{iwamoto_back_2009}
T.~Iwamoto, Y.~Sasayama, M.~Motoki, and T.~Kosaka, ``Back to mouth,'' in \emph{{{ACM SIGGRAPH}} 2009 {{Emerging Technologies}}}, 2009.

\bibitem{eamsa-ard_development_2018}
T.~{Eamsa-ard}, M.~M. Swe, T.~Seesaard, and T.~Kerdcharoen, ``Development of {{Electronic Nose}} for {{Evaluation}} of {{Fragrance}} and {{Human Body Odor}} in the {{Cosmetic Industry}},'' in \emph{2018 {{IEEE}} 7th {{Global Conference}} on {{Consumer Electronics}} ({{GCCE}})}.\hskip 1em plus 0.5em minus 0.4em\relax Nara, Japan: IEEE, Oct. 2018, pp. 363--364.

\bibitem{roddiger_vocnea_2018}
T.~R{\"o}ddiger, M.~Beigl, M.~K{\"o}pke, and M.~Budde, ``{{VOCNEA}}: Sleep apnea and hypopnea detection using a novel tiny gas sensor,'' in \emph{Proceedings of the 2018 {{ACM International Symposium}} on {{Wearable Computers}}}.\hskip 1em plus 0.5em minus 0.4em\relax Singapore Singapore: ACM, Oct. 2018, pp. 226--227.

\bibitem{chwalek_airspec_2023}
P.~Chwalek, S.~Zhong, D.~Ramsay, N.~Perry, and J.~Paradiso, ``{{AirSpec}}: {{A Smart Glasses Platform}}, {{Tailored}} for {{Research}} in the {{Built Environment}},'' in \emph{Adjunct {{Proceedings}} of the 2023 {{ACM International Joint Conference}} on {{Pervasive}} and {{Ubiquitous Computing}} \& the 2023 {{ACM International Symposium}} on {{Wearable Computing}}}.\hskip 1em plus 0.5em minus 0.4em\relax Cancun, Quintana Roo Mexico: ACM, Oct. 2023, pp. 204--204.

\bibitem{dam_wearable_2017}
N.~Dam, A.~Ricketts, B.~Catlett, and J.~Henriques, ``Wearable sensors for analyzing personal exposure to air pollution,'' in \emph{2017 {{Systems}} and {{Information Engineering Design Symposium}} ({{SIEDS}})}.\hskip 1em plus 0.5em minus 0.4em\relax Charlottesville, VA, USA: IEEE, Apr. 2017, pp. 1--4.

\bibitem{maag_w-air_2018}
B.~Maag, Z.~Zhou, and L.~Thiele, ``W-{{Air}}: {{Enabling Personal Air Pollution Monitoring}} on {{Wearables}},'' \emph{Proceedings of the ACM on Interactive, Mobile, Wearable and Ubiquitous Technologies}, vol.~2, no.~1, pp. 1--25, Mar. 2018.

\bibitem{geczy_wearable_2020}
A.~Geczy, L.~Kuglics, L.~Jakab, and G.~Harsanyi, ``Wearable {{Smart Prototype}} for {{Personal Air Quality Monitoring}},'' in \emph{2020 {{IEEE}} 26th {{International Symposium}} for {{Design}} and {{Technology}} in {{Electronic Packaging}} ({{SIITME}})}.\hskip 1em plus 0.5em minus 0.4em\relax Pitesti, Romania: IEEE, Oct. 2020, pp. 84--88.

\bibitem{zhong_hilo-wear_2020}
S.~Zhong, H.~S. Alavi, and D.~Lalanne, ``Hilo-wear: {{Exploring Wearable Interaction}} with {{Indoor Air Quality Forecast}},'' in \emph{Extended {{Abstracts}} of the 2020 {{CHI Conference}} on {{Human Factors}} in {{Computing Systems}}}.\hskip 1em plus 0.5em minus 0.4em\relax Honolulu HI USA: ACM, Apr. 2020, pp. 1--8.

\bibitem{palomeque-mangut_wearable_2022}
S.~{Palomeque-Mangut}, F.~Mel{\'e}ndez, J.~{G{\'o}mez-Su{\'a}rez}, S.~{Frutos-Puerto}, P.~Arroyo, E.~{Pinilla-Gil}, and J.~Lozano, ``Wearable system for outdoor air quality monitoring in a {{WSN}} with cloud computing: {{Design}}, validation and deployment,'' \emph{Chemosphere}, vol. 307, p. 135948, Nov. 2022.

\bibitem{roberts_decoding_2020}
S.~C. Roberts, P.~K. Misztal, and B.~Langford, ``Decoding the social volatilome by tracking rapid context-dependent odour change,'' \emph{Philosophical Transactions of the Royal Society B: Biological Sciences}, vol. 375, no. 1800, p. 20190259, Jun. 2020.

\bibitem{takahashi_feasibility_2008}
K.~Takahashi and I.~Sugimoto, ``Feasibility of emotion recognition from breath gas information,'' in \emph{2008 {{IEEE}}/{{ASME International Conference}} on {{Advanced Intelligent Mechatronics}}}.\hskip 1em plus 0.5em minus 0.4em\relax Xian, China: IEEE, Jul. 2008, pp. 625--630.

\bibitem{takahashi_remarks_2009}
------, ``Remarks on emotion recognition from breath gas information,'' in \emph{2009 {{IEEE International Conference}} on {{Robotics}} and {{Biomimetics}} ({{ROBIO}})}.\hskip 1em plus 0.5em minus 0.4em\relax Guilin, China: IEEE, Dec. 2009, pp. 938--943.

\bibitem{wicker_cinema_2015}
J.~Wicker, N.~Krauter, B.~Derstorff, C.~St{\"o}nner, E.~Bourtsoukidis, T.~Kl{\"u}pfel, J.~Williams, and S.~Kramer, ``Cinema {{Data Mining}}: {{The Smell}} of {{Fear}},'' in \emph{Proceedings of the 21th {{ACM SIGKDD International Conference}} on {{Knowledge Discovery}} and {{Data Mining}}}.\hskip 1em plus 0.5em minus 0.4em\relax Sydney NSW Australia: ACM, Aug. 2015, pp. 1295--1304.

\bibitem{williams_cinema_2016}
J.~Williams, C.~St{\"o}nner, J.~Wicker, N.~Krauter, B.~Derstroff, E.~Bourtsoukidis, T.~Kl{\"u}pfel, and S.~Kramer, ``Cinema audiences reproducibly vary the chemical composition of air during films, by broadcasting scene specific emissions on breath,'' \emph{Scientific Reports}, vol.~6, no.~1, p. 25464, May 2016.

\bibitem{bensemann_what_2023}
J.~Bensemann, H.~Cheena, D.~T.~J. Huang, E.~Broadbent, J.~Williams, and J.~Wicker, ``From {{What You See}} to {{What We Smell}}: {{Linking Human Emotions}} to {{Bio-markers}} in {{Breath}},'' \emph{IEEE Transactions on Affective Computing}, pp. 1--13, 2023.

\bibitem{wei_one-year_2023}
J.~Wei, Y.~Wang, J.~Mo, and C.~Fan, ``One-year dataset of hourly air quality parameters from 100 air purifiers used in {{China}} residential buildings,'' \emph{Scientific Data}, vol.~10, no.~1, p. 715, Oct. 2023.

\bibitem{betancourt_aq-bench_2021}
C.~Betancourt, T.~Stomberg, R.~Roscher, M.~G. Schultz, and S.~Stadtler, ``{{AQ-Bench}}: A benchmark dataset for machine learning on global air quality metrics,'' \emph{Earth System Science Data}, vol.~13, no.~6, pp. 3013--3033, Jun. 2021.

\bibitem{barcelo-ordinas_h2020_2021}
J.~M. {Barcelo-Ordinas}, P.~{Ferrer-Cid}, J.~{Garcia-Vidal}, M.~Viana, and A.~Ripoll, ``H2020 project {{CAPTOR}} dataset: {{Raw}} data collected by low-cost {{MOX}} ozone sensors in a real air pollution monitoring network,'' \emph{Data in Brief}, vol.~36, p. 107127, Jun. 2021.

\bibitem{ferrer-cid_raw_2022}
P.~{Ferrer-Cid}, J.~M. {Barcelo-Ordinas}, and J.~{Garcia-Vidal}, ``Raw data collected from {{NO2}} , {{O3}} and {{NO}} air pollution electrochemical low-cost sensors,'' \emph{Data in Brief}, vol.~45, p. 108586, Dec. 2022.

\bibitem{rodriguez_gamboa_electronic_2019-1}
J.~C. Rodriguez~Gamboa, E.~S. Albarracin~E., A.~J. Da~Silva, and T.~A. E.~Ferreira, ``Electronic nose dataset for detection of wine spoilage thresholds,'' \emph{Data in Brief}, vol.~25, p. 104202, Aug. 2019.

\bibitem{wijaya_electronic_2018}
D.~R. Wijaya, R.~Sarno, and E.~Zulaika, ``Electronic nose dataset for beef quality monitoring in uncontrolled ambient conditions,'' \emph{Data in Brief}, vol.~21, pp. 2414--2420, Dec. 2018.

\bibitem{erlangga_electronic_2021}
F.~Erlangga, D.~R. Wijaya, and W.~Wikusna, ``Electronic {{Nose Dataset}} for {{Classifying Rice Quality}} using {{Neural Network}},'' in \emph{2021 9th {{International Conference}} on {{Information}} and {{Communication Technology}} ({{ICoICT}})}.\hskip 1em plus 0.5em minus 0.4em\relax Yogyakarta, Indonesia: IEEE, Aug. 2021, pp. 462--466.

\bibitem{wijaya_electronic_2022}
D.~R. Wijaya, R.~Sarno, E.~Zulaika, and F.~Afianti, ``Electronic nose homogeneous data sets for beef quality classification and microbial population prediction,'' \emph{BMC Research Notes}, vol.~15, no.~1, p. 237, Dec. 2022.

\bibitem{sarno_electronic_2020-1}
R.~Sarno, S.~I. Sabilla, D.~R. Wijaya, D.~Sunaryono, and C.~Fatichah, ``Electronic nose dataset for pork adulteration in beef,'' \emph{Data in Brief}, vol.~32, p. 106139, Oct. 2020.

\bibitem{qu_open-set_2022}
C.~Qu, C.~Liu, Y.~Gu, S.~Chai, C.~Feng, and B.~Chen, ``Open-set gas recognition: {{A}} case-study based on an electronic nose dataset,'' \emph{Sensors and Actuators B: Chemical}, vol. 360, p. 131652, Jun. 2022.

\bibitem{fonollosa_chemical_2015}
J.~Fonollosa, I.~{Rodr{\'i}guez-Luj{\'a}n}, and R.~Huerta, ``Chemical gas sensor array dataset,'' \emph{Data in Brief}, vol.~3, pp. 85--89, Jun. 2015.

\bibitem{duran_acevedo_electronic_2021-1}
C.~M. Dur{\'a}n~Acevedo, C.~A. Cuastumal~Vasquez, and J.~K. Carrillo~G{\'o}mez, ``Electronic nose dataset for {{COPD}} detection from smokers and healthy people through exhaled breath analysis,'' \emph{Data in Brief}, vol.~35, p. 106767, Apr. 2021.

\bibitem{henderson_benchmarking_2020}
B.~Henderson, D.~M. Ruszkiewicz, M.~Wilkinson, J.~D. Beauchamp, S.~M. Cristescu, S.~J. Fowler, D.~Salman, F.~D. Francesco, G.~Koppen, J.~Langej{\"u}rgen, O.~Holz, A.~Hadjithekli, S.~Moreno, M.~Pedrotti, P.~Sinues, G.~Slingers, M.~Wilde, T.~Lomonaco, D.~Zanella, R.~Zenobi, J.-F. Focant, S.~{Grassin-Delyle}, F.~A. Franchina, M.~Mal{\'a}skov{\'a}, P.-H. Stefanuto, G.~Pugliese, C.~Mayhew, and C.~L.~P. Thomas, ``A benchmarking protocol for breath analysis: The peppermint experiment,'' \emph{Journal of Breath Research}, vol.~14, no.~4, p. 046008, Oct. 2020.

\bibitem{wilkinson_peppermint_2021}
M.~Wilkinson, I.~White, K.~Hamshere, O.~Holz, S.~Schuchardt, F.~G. Bellagambi, T.~Lomonaco, D.~Biagini, F.~F. Di, and S.~J. Fowler, ``The peppermint breath test: A benchmarking protocol for breath sampling and analysis using {{GC}}--{{MS}},'' \emph{Journal of Breath Research}, vol.~15, no.~2, p. 026006, Apr. 2021.

\bibitem{rottenberg_emotion_2007}
J.~Rottenberg, R.~Ray, and J.~Gross, ``Emotion elicitation using films,'' in \emph{The {{Handbook}} of {{Emotion Elicitation}} and {{Assessment}}}.\hskip 1em plus 0.5em minus 0.4em\relax New York, New York, USA: Oxford University Press, Mar. 2007.

\bibitem{schaefer_assessing_2010}
A.~Schaefer, F.~Nils, X.~Sanchez, and P.~Philippot, ``Assessing the effectiveness of a large database of emotion-eliciting films: {{A}} new tool for emotion researchers,'' \emph{Cognition \& Emotion}, vol.~24, no.~7, pp. 1153--1172, Nov. 2010.

\bibitem{saganowski_emognition_2022}
S.~Saganowski, J.~Komoszy{\'n}ska, M.~Behnke, B.~Perz, D.~Kunc, B.~Klich, {\L}.~D. Kaczmarek, and P.~Kazienko, ``Emognition dataset: Emotion recognition with self-reports, facial expressions, and physiology using wearables,'' \emph{Scientific Data}, vol.~9, no.~1, p. 158, Apr. 2022.

\bibitem{benz_nature-based_2022}
A.~B.~E. Benz, R.~J. Gaertner, M.~Meier, E.~Unternaehrer, S.~Scharndke, C.~Jupe, M.~Wenzel, U.~U. Bentele, S.~J. Dimitroff, B.~F. Denk, and J.~C. Pruessner, ``Nature-{{Based Relaxation Videos}} and {{Their Effect}} on {{Heart Rate Variability}},'' \emph{Frontiers in Psychology}, vol.~13, p. 866682, Jun. 2022.

\bibitem{soleymani_multimodal_2012}
M.~Soleymani, J.~Lichtenauer, T.~Pun, and M.~Pantic, ``A {{Multimodal Database}} for {{Affect Recognition}} and {{Implicit Tagging}},'' \emph{IEEE Transactions on Affective Computing}, vol.~3, no.~1, pp. 42--55, Jan. 2012.

\bibitem{li_recognizing_2020}
W.~Li, Z.~Jia, D.~Xie, K.~Chen, J.~Cui, and H.~Liu, ``Recognizing lung cancer using a homemade e-nose: {{A}} comprehensive study,'' \emph{Computers in Biology and Medicine}, vol. 120, p. 103706, May 2020.

\bibitem{burnard_method_1991}
P.~Burnard, ``A method of analysing interview transcripts in qualitative research,'' \emph{Nurse Education Today}, vol.~11, no.~6, pp. 461--466, Dec. 1991.

\bibitem{maciel_optimization_2023}
M.~Maciel, S.~Sankari, M.~Woollam, and M.~Agarwal, ``Optimization of {{Metal Oxide Nanosensors}} and {{Development}} of a {{Feature Extraction Algorithm}} to {{Analyze VOC Profiles}} in {{Exhaled Breath}},'' \emph{IEEE Sensors Journal}, vol.~23, no.~15, pp. 16\,571--16\,578, Aug. 2023.

\bibitem{saidi_exhaled_2018}
T.~Saidi, O.~Zaim, M.~Moufid, N.~El~Bari, R.~Ionescu, and B.~Bouchikhi, ``Exhaled breath analysis using electronic nose and gas chromatography--mass spectrometry for non-invasive diagnosis of chronic kidney disease, diabetes mellitus and healthy subjects,'' \emph{Sensors and Actuators B: Chemical}, vol. 257, pp. 178--188, Mar. 2018.

\bibitem{voss_detection_2022}
A.~Voss, R.~Schroeder, S.~Schulz, J.~Haueisen, S.~Vogler, P.~Horn, A.~Stallmach, and P.~Reuken, ``Detection of {{Liver Dysfunction Using}} a {{Wearable Electronic Nose System Based}} on {{Semiconductor Metal Oxide Sensors}},'' \emph{Biosensors}, vol.~12, no.~2, p.~70, Jan. 2022.

\bibitem{haddad_feature_2007}
R.~Haddad, L.~Carmel, and D.~Harel, ``A feature extraction algorithm for multi-peak signals in electronic noses,'' \emph{Sensors and Actuators B: Chemical}, vol. 120, no.~2, pp. 467--472, Jan. 2007.

\bibitem{shakya_high-dimensional_2020}
P.~Shakya, E.~Kennedy, C.~Rose, and J.~K. Rosenstein, ``High-{{Dimensional Time Series Feature Extraction}} for {{Low-Cost Machine Olfaction}},'' \emph{IEEE Sensors Journal}, pp. 1--1, 2020.

\bibitem{xian-min_detecting_2013}
M.~{Xian-Min}, ``Detecting of {{Coal Gas Weak Signals Using Lyapunov Exponent}} under {{Strong Noise Background}},'' in \emph{2013 {{Third International Conference}} on {{Intelligent System Design}} and {{Engineering Applications}}}.\hskip 1em plus 0.5em minus 0.4em\relax China, Hong Kong: IEEE, Jan. 2013, pp. 583--586.

\bibitem{sabilla_development_2017}
S.~I. Sabilla and R.~Sarno, ``Development of wavelet transforms to predict methane in chili using the electronic nose,'' in \emph{2017 {{International Conference}} on {{Advanced Mechatronics}}, {{Intelligent Manufacture}}, and {{Industrial Automation}} ({{ICAMIMIA}})}.\hskip 1em plus 0.5em minus 0.4em\relax Surabaya: IEEE, Oct. 2017, pp. 271--276.

\bibitem{schmuker_exploiting_2016}
M.~Schmuker, V.~Bahr, and R.~Huerta, ``Exploiting plume structure to decode gas source distance using metal-oxide gas sensors,'' \emph{Sensors and Actuators B: Chemical}, vol. 235, pp. 636--646, Nov. 2016.

\bibitem{burgues_smelling_2019}
J.~Burgu{\'e}s, V.~Hern{\'a}ndez, A.~Lilienthal, and S.~Marco, ``Smelling {{Nano Aerial Vehicle}} for {{Gas Source Localization}} and {{Mapping}},'' \emph{Sensors}, vol.~19, no.~3, p. 478, Jan. 2019.

\end{thebibliography}
\appendix

\begin{table*}[htb]
\caption{Summary of dataset's key files.}
\begin{center}
\scriptsize
\begin{tabularx}{\linewidth}{lX}
\toprule
\textbf{File} & \textbf{Description}                                      \\ \midrule
\rowcolor[HTML]{EFEFEF} 
\path{AAQ_Dataset/AAQ_Qualtrics_Intake.csv}           & Responses from all participants to the intake form. \\
\path{AAQ_Dataset/Gas_Sensor_Data/<P>_Gas-Data_Raw.csv}           & Contains the recorded gas sensor readings from the headset and desktop monitors, where \path{<P>} is the participant number. \\
\rowcolor[HTML]{EFEFEF} 
\path{AAQ_Dataset/Gas_Sensor_Data/<P>_EmotionRatings_<DATE>.csv}       & Contains the valence, arousal, and familiarity ratings alongside timestamps for all portions of study (e.g., video start and end times) to segment the raw data, where \path{<P>} is the participant number and \path{<DATE>} is the date of data collection.               \\ 
\path{AAQ_Dataset/Participant_Logs.csv}    & Summary of participant data collection date, status, and sampling frequency. \\
\rowcolor[HTML]{EFEFEF} 
\path{AAQ_Dataset/Interview_Data/<P>_InterviewTranscript.txt} & Each exit interview transcript, where \path{<P>} denotes the participant number.\\
\bottomrule
\end{tabularx}
\label{tab:files}
\end{center}
\end{table*}

\subsection{Review of Established Gas Sensor Features}
As we recorded continuous exposure of the gas sensors to breathing, our feature extraction process did not match the sampling conditions of previous research on electronic nose research or breath analysis. Following, we first collated features from sub-fields of study. Two major differences in sampling and feature extraction emerged.

First, several papers focused on extracting features from either a single exhalation \cite{li_recognizing_2020, maciel_optimization_2023, saidi_exhaled_2018}, respiration cycle \cite{voss_detection_2022}, or sample injection \cite{haddad_feature_2007, shakya_high-dimensional_2020}, which we found to be -- overall -- similar processes where exposure of the MOX gas sensors to the target gas is controlled and limited to encompass both the response and return to baseline (covering both the adsorption and absorption phases). Relevant features mentioned included typical statistical features but also peak-to-peak values, the ratio of one sensor's signal to total response of all sensors (``relative abundance'' \cite{maciel_optimization_2023}), max conductance change from baseline, R\'enyi entropy \cite{voss_detection_2022}, autocorrelation, automutual information, wavelet entropy, the Generalized Lorentzian \cite{haddad_feature_2007}, amongst others \cite{shakya_high-dimensional_2020}.

Second are papers focused on continuous monitoring or exposure to gases \cite{xian-min_detecting_2013, sabilla_development_2017, schmuker_exploiting_2016, burgues_smelling_2019}, where -- often -- the gas sensor signals are either continuously exposed to a changing gas mixture or under a turbulent environment (i.e., using gas sensing robots to locate a gas source). Alongside standard statistical features, papers proposed features such as the Lyapunov exponent \cite{xian-min_detecting_2013}, Discrete Wavelet Transform (DWT) coefficients \cite{sabilla_development_2017}, and a feature based on the transient response of MOX gas sensor signals (``bouts'' \cite{schmuker_exploiting_2016, burgues_smelling_2019}). These features tend to assume that the gas sensors may not reach a steady state, as the gas conditions may fluctuate constantly.

\end{document}